\documentclass[useAMS,usenatbib,usegraphicx]{mn2e}

   \def\aap{A\&A}
\def\apjl{ApJ} \def\apj{ApJ} \def\apjs{ApJS} \def\aj{AJ} \def\mnras{MNRAS}
 \def\pasp{PASP}  \def\procspie{Proceedings of the SPIE}

\title[Massive Double White Dwarfs]{Massive Double White Dwarfs and the AM CVn Birthrate}

\author[M. Kilic et al.]
{
Mukremin Kilic$^{1}$,
Warren R. Brown$^2$,
Craig O. Heinke$^3$,
A. Gianninas$^{1}$,
P. Benni$^4$,
\newauthor
and M. A. Ag\"{u}eros$^5$\\
$^1$Homer L. Dodge Department of Physics and Astronomy, University of Oklahoma,
440 W. Brooks St., Norman, OK, 73019, USA\\
$^2$Smithsonian Astrophysical Observatory, 60 Garden St, Cambridge, MA 02138, USA\\
$^3$Department of Physics, CCIS 4-183, University of Alberta, Edmonton, AB T6G 2E1, Canada\\
$^4$Acton Sky Portal, 3 Concetta Circle, Acton, MA 01720, USA\\
$^5$Department of Astronomy, Columbia University, 550 West 120th Street, New York, NY 10027, USA
}

\begin{document}

\maketitle

\begin{abstract}

We present Chandra and Swift X-ray observations of four extremely low-mass (ELM) white dwarfs
with massive companions. We place stringent limits on X-ray emission from all four systems,
indicating that neutron star companions are extremely unlikely and that the companions are
almost certainly white dwarfs. Given the observed orbital
periods and radial velocity amplitudes, the total masses of these binaries are greater than
1.02 to 1.39 M$_{\odot}$. The extreme mass ratios between the two components make it unlikely
that these binary white dwarfs will merge and explode as Type Ia or underluminous supernovae.
Instead, they will likely
go through stable mass transfer through an accretion disk and turn into interacting AM CVn.
Along with three previously known systems, we identify two of our targets, J0811 and J2132, as
systems that will definitely undergo stable mass transfer. In addition, we use the binary
white dwarf sample from
the ELM Survey to constrain the inspiral rate of systems with extreme mass ratios. This rate,
$1.7 \times10^{-4}$ yr$^{-1}$, is consistent with the AM CVn space density estimated from the
Sloan Digital Sky Survey. Hence, stable mass transfer double white dwarf progenitors can account
for the entire AM CVn population in the Galaxy.

\end{abstract}

\begin{keywords}
binaries: close ---
white dwarfs ---
stars: individual (SDSS J075519.47+480034.0, 
SDSS J081133.56+022556.7, 
SDSS J144342.74+150938.6,
SDSS J213228.36+075428.2)
\end{keywords}

\section{INTRODUCTION}

Short period binary white dwarfs will lose angular momentum through gravitational wave radiation and start mass
transfer. The shortest period detached double white dwarf system currently known, J0651 \citep{brown11,hermes12}, will
start mass transfer in less than 1 Myr. What happens next depends on the stability of mass transfer. 

For roughly equal mass binaries, the two stars will merge and form a more massive white dwarf (for CO+CO white dwarfs), an
R Cor Bor star (for CO+He white dwarf mergers), or a single subdwarf (for He+He white dwarfs). Depending on the total mass of
the system, the mergers can also lead to underluminous or normal Type Ia supernovae in both CO+CO and
CO+He white dwarf mergers \citep{webbink84,iben84}. The latter systems may go through a double detonation, in which the detonation of the
surface He layer leads to the detonation of the underlying CO core white dwarf \citep[e.g.,][]{shen14}.

For binaries with extreme mass ratios, the mass transfer will be stable \citep{marsh04}, leading to the formation of
AM CVn binaries. AM CVn have orbital periods of 5-65 min and involve an accreting white dwarf and a He-rich donor star.
Given the stability of the mass transfer, the orbit slowly expands to accommodate the increasing size of the
degenerate donor, the mass transfer rate decreases, and the binary evolves into a massive white dwarf with a
planetary size companion. Despite large scale efforts to find AM CVn \citep{carter13},
there are only 52 systems currently known. Hence, AM CVn represent the end product of a rare and
fine-tuned evolution in binary systems \citep{solheim10}.

There are three formation channels for AM CVn with three different donor stars; low-mass white dwarfs,
helium stars, or evolved main-sequence stars. 
The population synthesis models predict that the double white dwarf channel
dominates the AM CVn formation \citep{nelemans01,nissanke12}. However,
it is difficult to distinguish between the different
formation channels based on the observed AM CVn population, since all three channels essentially
lead to very low-mass degenerate He-rich donors. \citet{nelemans10} demonstrate that N/C abundance ratios 
significantly differ between the He-white dwarf and He star donors, and use this to identify He white dwarf donors
in three AM CVn systems. \citet{breedt12} present the first compelling evidence for an AM CVn progenitor
with an evolved main-sequence donor. They constrain the orbital period of CSS1122-1110, a cataclysmic
variable 
with unusually strong He lines, to 65.2 min and use the superhump period excess to infer a mass ratio of 0.017.
\citet{kennedy15} identify three other systems with evolved main-sequence donors as potential progenitors of
AM CVn, though the mass ratios are not as extreme as CSS1122-1110.
Observationally, it remains unclear what is the dominant formation channel for AM CVn systems.

The ELM Survey \citep[][and references therein]{brown16a} has found 76 short period double white dwarfs so far,
including several systems with extreme mass ratios. \citet{kilic14} identify two of these systems,
J0751-0141 and J1741+6526, as the first confirmed AM CVn progenitors from the double white dwarf channel. Here we present
follow-up observations of four more ELM white dwarfs with massive companions, and demonstrate that two of these targets
will have stable mass transfer. In addition, we estimate the merger rate of ELM white dwarfs with massive companions.
We present our target
selection and observations in \S 2, and discuss the nature of the companions
and the inspiral rate of extreme mass ratio binaries in the ELM Survey in \S 3. We conclude
in \S 4.

\section{Observations and Results}

\subsection{Target Selection and Motivation for X-ray Observations}

ELM white dwarfs are single-lined spectroscopic binaries in which the ELM white dwarf dominates
the light of the system. We define $M_1$ as the visible low mass white dwarf and $M_2$ as its unseen companion.
We measure orbital period and radial velocity semi-amplitude, and
derive $M_1$ by comparing our spectroscopic $\log{g}$ and $T_{\rm eff}$ measurements to
evolutionary tracks for low-mass He-core white dwarfs \citep{althaus13}. In the absence of information about inclination, such
as eclipses, the observations provide a lower limit on $M_2$. 

\begin{figure}
\includegraphics[width=3.5in,angle=0]{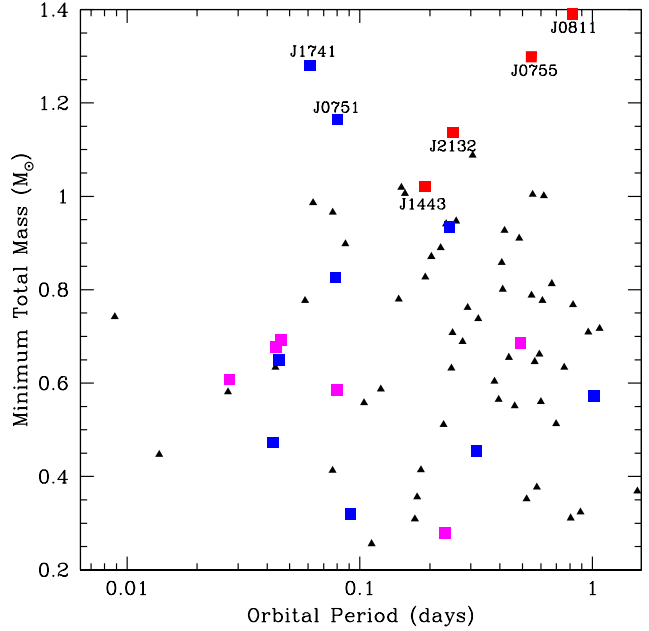}
\caption{The minimum total system mass versus binary orbital period
for the ELM Survey VII sample \citep{brown16a}. Blue and magenta squares mark the objects with
previous X-ray and radio observations, respectively. The two previously confirmed AM CVn progenitor systems,
J0751 and J1741, as well as the four targets with new X-ray observations (red squares) presented
in this paper are labeled.}
\label{fig:target}
\end{figure}

Figure \ref{fig:target} shows the minimum total system mass versus the orbital period
for the 76 binaries discovered in the ELM Survey. Nine of these targets have previous Chandra
or XMM-Newton observations that rule out neutron star companions, and an additional six targets
have radio data that rule out milli-second pulsars. To search for neutron star companions, we
selected four of the most massive binary systems known from an earlier version of the ELM Survey
sample for follow-up X-ray observations. Table 1 presents the physical parameters of these four
systems, including the minimum companion masses \citep{gianninas15}.

\begin{table*}
\centering
\caption{Physical parameters of our targets.}
\begin{tabular}{lccccccc}
\hline
{SDSS} & {$T_{\rm eff}$} & {$\log{g}$} & {$M_1$} & {$R_1$} & {Period} & {$M_{2,min}$} & {$a$} \\
       &  (K) & (cm s$^{-2}$) & ($M_{\odot}$) & ($R_{\odot}$) & (days) & ($M_{\odot}$) & ($R_{\odot}$)\\
\hline
J075519.47+480034.0 & 19530 $\pm$ 300 & 7.42 $\pm$ 0.05 & 0.41 & 0.0207 & 0.54627 & 0.89 & 3.07 \\
J081133.56+022556.7 & 13540 $\pm$ 200 & 5.67 $\pm$ 0.05 & 0.18 & 0.1035 & 0.82194 & 1.21 & 4.12 \\
J144342.74+150938.6 &  8970 $\pm$ 130 & 6.44 $\pm$ 0.06 & 0.18 & 0.0424 & 0.19053 & 0.84 & 1.40 \\
J213228.36+075428.2 & 13790 $\pm$ 200 & 6.02 $\pm$ 0.04 & 0.18 & 0.0681 & 0.25056 & 0.96 & 1.75 \\
\hline
\end{tabular}
\end{table*}

Given the short orbital periods ($P\leq1$ d) and very low masses for the observed white dwarfs (requiring
evolutionary stripping), neutron star companions to our targets would
be spun up to millisecond periods. Such millisecond pulsars can be detected in the radio,
but the radio pulsar beam may miss our line of sight. Instead, X-ray observations enable
us to observe the blackbody emission from $>75$\% of the neutron star surface. 
\citet{heinke05} and \citet{bogdanov06} detected all 15 radio millisecond pulsars in
unconfused regions of 47 Tuc in X-rays, with $L_X(0.5-6 keV) \geq 2\times10^{30}$ erg s$^{-1}$.
Continued measurements of X-ray fluxes and distances of radio millisecond pulsars have revealed
a few systems with lower X-ray luminosities \citep[e.g.,][]{pavlov07,kargaltsev12,forestell14,spiewak16}.
Of 52 millisecond pulsars in the list of \citet{forestell14}, only four have X-ray luminosities
below $10^{30}$ erg/s \footnote{\citet{desvignes16} provide a parallax distance for PSR
J1024-0719 of 1083$^{+226}_{-163}$ pc, indicating $L_X=2.8\times10^{30}$ erg/s.}. Hence, deep X-ray
observations that place an $L_X$ limit at or below $10^{30}$ erg/s can confirm or provide strong
evidence against neutron star companions to our targets.

\subsection{No Neutron Star Companions}

We observed three targets, J0811, J1443, and J2132, with Chandra's ACIS-S detector in Very Faint mode, for 32.6,
1.3, and 11.6 ks, respectively. These observations were performed between 2014 Dec and 2015 Feb. 
We reprocessed the raw Chandra data using {\sevensize CIAO 4.5}. Inspection of the 0.3-7 keV images
revealed only 1 photon within $1\arcsec$ of the position of J0811, and zero photons for the other two white dwarfs.

We also obtained 1.1~ks Swift XRT observations of J0755 on UT 2014 March 10. Unfortunately, J0755 landed
on a bad column in the XRT, which reduced the effective exposure time by a factor of three.
We obtained an additional 1.0~ks Swift observation of the same target on UT 2016 Feb 22.  
We used a $20\arcsec$ region for X-ray extraction, which encloses 80\% of the Swift/XRT point spread function
\citep{moretti04}. We find zero photons within this extraction region.

We use the {\sevensize COLDEN} tool to interpolate the \citet{dickey90} HI survey, and estimate $N_H$ values
for each target. We also use the {\sevensize PIMMS} tool to compute the unabsorbed 0.3-8 keV flux for a 134
eV blackbody \citep[appropriate for the faintest 47 Tuc millisecond pulsar, 47 Tuc T,][]{bogdanov06}. We calculate 95\% confidence
upper limits \citep[4.7 counts for J0811 and 3 counts for the other three stars,][]{gehrels86},
to the 0.3-8 keV X-ray luminosities. Table 2 presents these upper limits for each target. 

The 95\% confidence upper $L_X$ limits for all four targets are a factor of two or more
below the luminosity of any millisecond pulsar in 47 Tuc, using the blackbody fluxes reported
in \citet{bogdanov06} and a 4.5 kpc distance \citep[][2010 revision]{harris96}.
We conclude that our X-ray observations provide strong evidence against neutron star companions
for all four targets.

Radio and optical follow-up observations of pulsars indicate that He white dwarf companions 
are common. There are more than 100 pulsar + He white dwarf systems known in the Australia
Telescope National Facility Pulsar Catalogue \citep[2016 version,][]{manchester05}. However,
the reverse approach, the search for milli-second pulsar companions through X-ray or radio follow-up
of ELM white dwarfs, has so far resulted in no new pulsar discoveries
\citep[][and this study]{vanleeuwen07,agueros09,kilic11,kilic12,kilic14}. Based on
a statistical analysis of the companion mass distribution to ELM white dwarfs, \citet{andrews14}
find a neutron star companion fraction of $<$16\% \citep[see also][]{boffin15,brown16a}. Hence,
follow-up X-ray or radio observations of a larger sample of ELM white dwarfs are necessary
to discover the first pulsar through its white dwarf companion.

\begin{table*}
\centering
\caption{The X-ray limits on our targets.}
\begin{tabular}{lcccrcc}
\hline
{Name} & {ObsID} & {Dist} & {$N_H$}     & {Exp} &  {Count rate}    & {$L_X$} \\
       &         &  (kpc) & (cm$^{-2}$) &   (ks)  &  (cts s$^{-1}$) & (ergs s$^{-1}$) \\
\hline
J0755 & 33183002 & 0.17 & $4.8\times 10^{20}$ &  1.0  &  $<3.8\times 10^{-3}$ & $<5.6\times 10^{29}$ \\
J0811 & 16687    & 2.20 & $3.5\times 10^{20}$ & 32.6  &  $<1.4\times 10^{-4}$ & $<1.1\times 10^{30}$ \\
J1443 & 16685    & 0.54 & $1.5\times 10^{20}$ &  1.3  &  $<2.3\times 10^{-3}$ & $<1.0\times 10^{30}$ \\
J2132 & 16686    & 1.09 & $4.4\times 10^{20}$ & 11.6  &  $<2.6\times 10^{-4}$ & $<5.4\times 10^{29}$ \\
\hline
\end{tabular}
\end{table*}

\subsection{Optical Photometry}

A significant number of short period binary white dwarfs display photometric variations due to
tidal distortions, the relativistic beaming effect, or eclipses \citep{hermes14}. The photometric
variations are a function of the white dwarf radii, orbital separation, and orbital inclination.
For example, the two previously
confirmed AM CVn progenitors, J0751 and J1741 display 3.2\% and 1.3\% ellipsoidal variations.
The amplitude of the ellipsoidal effect is roughly $\Delta f_{ell} = (m_2/m_1)(r_1/a)^3$, where
$a$ is the orbital semi-major axis and $r_1$ is the radius of the primary \citep{zucker07,shporer10}.
The four targets presented in Table 1 have orbital periods and separations significantly larger
than J0751 and J1741. Hence, the expected amplitude of the ellipsoidal effect
is smaller than 0.05\% for all four targets. 

\begin{figure*}
\includegraphics[width=3.5in,angle=-90]{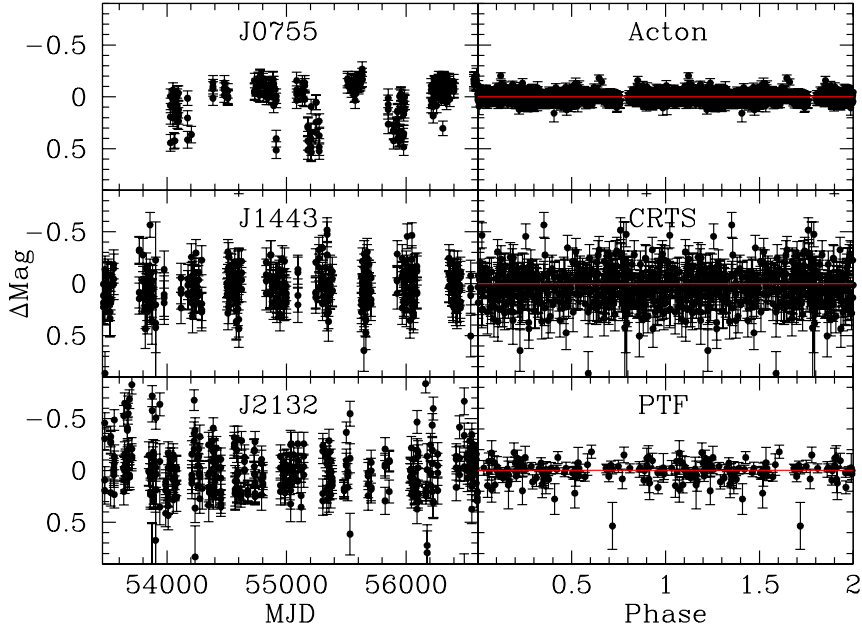}
\caption{{\it Left Panels:} Catalina Sky Survey light curves for three of our targets.
{\it Right Panels:} Phase folded light curves (using the orbital period measured from the
radial velocity data) for the same three targets based on the Acton
Sky Portal (top), Catalina Sky Survey (middle), and Palomar Transient Factory (bottom) data.}
\label{fig:lc}
\end{figure*}

Figure \ref{fig:lc} displays the unfiltered optical light curves for three of our targets from the Catalina Sky Survey \citep{drake09}. The remaining target, J0811, has a galaxy within
6.6$\arcsec$, and its light curve is not available in the Catalina Sky Survey Data Release 2.
J0811 was not observed by the Palomar Transient Factory \citep[PTF,][]{rau09}. Hence, we do not
have any photometric constraints on this binary.
The Catalina data for J0755, J1143, and J2132 are relatively noisy, and a Fourier analysis does not reveal any significant periodicities for the latter two systems.
The right panels show the Catalina and $R-$band PTF light curves for
these two stars, folded on the best-fit period from the radial velocity data. There are no significant
photometric variations at the orbital period for J1143 and J2132.

The Catalina data for J0755 show large scatter that is consistent with long term
variations of order $P>600$ d. However, J0755 is within 18$\arcsec$ of a $g=13.88$ mag star, and it is not
clear if the Catalina photometry is affected by this nearby source. We obtained follow-up $V$-band optical
photometry of J0755 over 20.7 h between UT 2016 Feb 28 and Mar 3 using Celestron 28-cm and 35-cm
Schmidt-Cassegrain telescopes at the Acton Sky Portal private observatory. The top right panel in
Fig. \ref{fig:lc} shows the phase-folded light curve for J0755. There is no evidence of photometric
variations at the orbital period for J0755 in our data, as well as the PTF data that covers
MJD 55081-57468 (T. Kupfer 2016, priv. comm). Given the precision of our light curves and the
relatively small amplitudes of the predicted ellipsoidal variations and the Doppler beaming effect,
the absence of optical photometric variations in J0755, as well as J1443 and J2132, is not surprising.
Unfortunately, the absence of photometric variations leaves us with no additional information on the binaries.

\section{Discussion}

\subsection{Four Massive Double White Dwarfs}

Our Chandra and Swift X-ray observations demonstrate that all four of our low-mass white dwarf targets
almost certainly have massive white dwarf companions. J0811 is the most extreme system in our sample 
with a 0.18 $M_{\odot}$ white dwarf 
and a $\geq1.21 M_{\odot}$ white dwarf companion. However, the merger time due to gravitational wave radiation
is longer than a Hubble time. J2132 is the second most massive system, with
a 0.18 $M_{\odot}$ white dwarf and a $\geq 0.96 M_{\odot}$ white dwarf companion. Given the mass ratio of $q<0.2$, J2132
will start stable mass transfer in 5.5-7.2 Gyr and turn into an AM CVn. 

\begin{figure}
\hspace{-0.1in}
\includegraphics[width=2.7in,angle=-90]{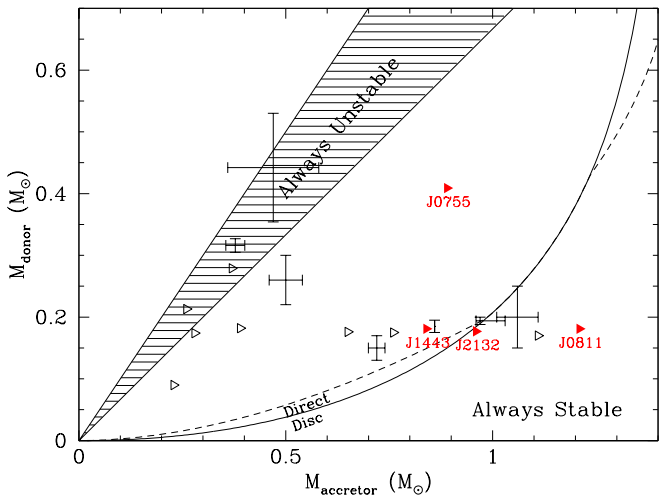}
\caption{Mass transfer stability for double white dwarfs \citep{marsh04}. Disk accretion occurs in
the region below the solid line. J0811 and J2132 are clearly in this parameter range.
Eclipsing double white dwarf systems and the double-lined binary J1257+5428 (error bars), and other
ELM white dwarf binaries with X-day data (triangles) are also shown.}
\label{fig:stab}
\end{figure}

Figure \ref{fig:stab} shows the mass transfer stability limits from \citet{marsh04}. Binaries
with a mass ratio of $q>\frac{2}{3}$ are expected to have unstable mass transfer and merge. The eclipsing,
short period double white dwarfs J1152+0248 \citep{hallakoun16}, CSS 41177 \citep{parsons11,bours14}, and the double lined binary
WD 1242-105 \citep{debes15} are excellent examples of systems in this region. Objects with mass ratios below
the solid line will have stable mass transfer, which leads to AM CVn systems instead of mergers.

Along with the previously identified AM CVn progenitors J0751, J1741 \citep{kilic14} and J1257+5428
\citep{kulkarni10,marsh11,bours15}, J0811 and J2132 are clearly in the stable mass transfer region \citep[see also][]{gianninas15}.
Out of these five systems, all but J0811 have merger times shorter than a Hubble time. Hence,
there are now four confirmed double white dwarf progenitors of AM CVn. 

In addition to these four systems, there are a number of other ELM white dwarfs
with unknown inclinations that are likely to have extreme mass ratios. For example, J1443 has a mass
ratio of $q\leq0.21$ (Fig. \ref{fig:stab}), which places it in the intermediate region where the stability
of mass transfer depends on the synchronization timescale (spin-orbit coupling) of the binary. 
Assuming that the orbital inclination is distributed randomly in $\sin{i}$ within the allowed
observational constraints, there is a 46\% chance that J1443 is in the disk accretion region.

\subsection{AM CVn Birthrate}

\citet{brown16a,brown16b} define a clean sample of 60 binary systems from the ELM Survey, which is
60\% complete in a well defined color and magnitude range. They identify 63\% of their sample as disk objects, and
use standard Galactic stellar density models to estimate the ELM white dwarf local space density. 
Using the stellar density model of \citet{nelemans01}, the merger rate
of disk ELM white dwarf binaries in the Milky Way is $5 \times10^{-3}$ yr$^{-1}$, with a factor of 1.8 uncertainty in this number.
\citet{carter13} find an AM CVn formation rate of $1.3 \pm 0.8 \times10^{-4}$ yr$^{-1}$, derived
using the same stellar density model and using observations from the same footprint of sky (SDSS).
Hence, the merger rate of all ELM white dwarf binaries is about 40 times larger than the AM CVn
formation rate. Based on this, \citet{brown16b} conclude that most double white dwarfs will not form AM CVn.

A remaining question is whether the inspiral rate of the extreme mass ratio systems in the ELM Survey
is consistent with the AM CVn formation rate. We perform Monte Carlo simulations of the clean, disk
ELM white dwarf binary sample to estimate the probability that each system has a mass ratio that puts it below
the stability limit of \citet{marsh04} for disk accretion. Since the ELM Survey binaries were selected
based on color, the distribution of inclination angles should be random. We include the constraints
on inclination for eclipsing systems, and the binaries with X-ray data. The allowed inclinations map
to an allowed distribution of $M_2$ for each binary.

Every ELM white dwarf binary has a likelihood of having mass ratios that fall in the stable mass transfer
region of parameter space, however not every binary will merge within a Hubble time. We estimate
the underlying merger rate by taking the observed distribution of merger times and assuming that
ELM white dwarf binaries form at a constant rate over the past Gyr, the approximate time span that ELM white dwarfs
are detectable in the ELM Survey color selection. 

The inspiral rate for the subset of binaries
with extreme mass ratios (that would lead to stable disk accretion) is a factor of 30 lower
than the total ELM white dwarf binary merger rate. This is due to factors of 10 longer gravitational
wave merger times for the extreme mass ratio systems in our sample, and factors of 3 fewer stars in
the stable mass transfer region. Hence, we estimate that the formation rate of stable mass transfer
systems from binary ELM white dwarfs is $1.7 \times10^{-4}$ yr$^{-1}$. This is remarkably similar to the AM CVn
formation rate found by \citet{carter13}. Note that both rates are uncertain by 60-80\%. Nevertheless,
the extreme mass ratio double white dwarf systems found in the ELM Survey can explain a significant fraction,
and perhaps the entire population, of AM CVn found in the Galaxy.

\section{Conclusions}

We provide strong evidence against neutron star companions in four ELM white dwarfs with massive companions.
J0755 and J0811 are the most massive binary white dwarfs identified in the ELM Survey, with total binary masses of 
$M \geq 1.3 M_{\odot}$ and $M \geq 1.39 M_{\odot}$, respectively. However, both J0755 and J0811 have orbital periods longer
than half a day. Hence, they will not interact within a Hubble time. The remaining two stars, J1443 and J2132,
have binary mass ratios of $q\leq0.21$, and they have short enough orbital periods to start mass transfer in
several Gyr. J2132 is clearly in the stable mass transfer range in Figure \ref{fig:stab}. Hence, there are
now four confirmed double white dwarf progenitors of AM CVn; J0751, J1741 \citep{kilic14},
J1257+5428 \citep{bours15}, and J2132. 

More importantly, we take the entire sample of ELM white dwarf binaries,
consider the distribution of mass ratios for each system, and derive the inspiral rate for the subset that
satisfy the \citet{marsh04} stability criterion for disk accretion. This rate is essentially
identical to the AM CVn formation rate from \citet{carter13};
there are sufficient numbers of double white dwarf progenitors of AM CVn to
account for the entire population of AM CVn in the Galaxy.

\section*{Acknowledgements}
We thank the organizers and attendees of the Fourth International Workshop on AM CVn Stars
for useful discussions.
Support for this work was provided by NASA through Chandra Award Number GO5-16022X issued
by the Chandra X-ray Observatory Center, which is operated by the Smithsonian Astrophysical 
Observatory for and on behalf of NASA under contract NAS8-03060.
Additional support was provided by the NSF and NASA under grants AST-1312678
and NNX14AF65G. COH was supported by an NSERC Discovery Grant.

\noindent {\it Facilities: CXO, Swift}

\end{document}